\documentclass[12pt,preprint]{aastex}

\begin{document}

\title {THE AGE-DEPENDENCE OF THE DETECTABILITY OF COMETS ORBITING SOLAR-TYPE STARS}

\author{M. Jura} 
\affil{Department of Physics and Astronomy, University of California,
    Los Angeles CA 90095-1562; jura@clotho.astro.ucla.edu}

\begin{abstract}
The  outflows from comets in orbit around G-type main-sequence stars can
be detected when they produce  transient OH absorption lines in the spectrum near 3100 {\AA} of the host star.  There is only about a  3 ${\times}$ 10$^{-8}$ probability of detecting an analog to comet  Hale-Bopp orbiting an analog to the Sun.  However,  for young  solar-type stars with very large numbers of  comets, possibly delivering water to terrestrial planets, there is as much as a 1\% chance that any sufficiently sensitive, 
randomly-timed observation may detect such transient absorption.

\end{abstract}

\section{INTRODUCTION}

A fundamental goal in modern astronomy is to learn more about the origin and evolution of the solar system and other planetary environments  where life may exist.    Circumstellar dust around main sequence
stars was discovered with IRAS (Aumann et al. 1984) and
has been studied extensively to learn indirectly about the origin and evolution of larger parent bodies  which might resemble asteroids or comets (Lagrange, Backman \& Artymowicz 2000; Zuckerman 2001).    To date, however, very little is directly known about these parent bodies.  The goal of this paper is
to describe an observational strategy to identify and characterize individual extrasolar
comets.

Water-rich asteroids and comets are probably responsible for delivering
to the Earth most of its water (Morbidelli et al. 2000).  By studying comets
and related objects around young stars, it may be possible to learn more
about this process which is fundamental for the development of life as we know it.  

There have been a number of previous discussions about possible observational signatures of comets
around other stars.  Alcock, Fristrom \& Siegelman (1986) suggested that the presence of metals in the atmospheres of white dwarfs could be explained by the collision of comets with the star, but this hypothesis
is not yet supported by  additional evidence (see Zuckerman et al. 2003). 
Melnick et al. (2001)  detected gas-phase H$_{2}$O in the outflow
from IRC+10216, a carbon-rich mass-losing red giant, which they interpreted
as being produced by the sublimation of comets (Stern, Shull \& Brandt 1990, Ford \& Neufeld 2001). However, the observed water might  result from chemical reactions in the dense outflow (Willacy 2004).  Jura (2004) has
argued that the lack of excess 25 ${\mu}$m radiation in IRAS data for first ascent red giants means that these stars typically have less than 0.1 M$_{\oplus}$ of   comet-like Kuiper 
Belt Objects in orbits at ${\sim}$ 45 AU.    

There is compelling evidence for Falling Evaporating Bodies (FEB's) around
${\beta}$ Pic, a 12 Myr-old star (Barrado y Navascues et al. 1999, Zuckerman et al. 2001) with a substantial amount of circumstellar dust (Artymowicz 1997, Vidal-Madjar, Lecavelier des Etangs \& Ferlet 1998). Transient absorption lines which can be attributed to infalling planetesimal
also have been detected around other young A-type stars (see, for example, Grady et al. 1996, Roberge et al. 2002). Although Lecavelier des Etangs, Vidal-Madjar \& Ferlet (1996) and Li \& Greenberg (1998) have proposed that much of
the dust around ${\beta}$ Pic arises from comets, the observed FEB's   contain a substantial amount of refractory material, and their degree of  resemblance to ice-rich comets in the solar system is  uncertain (Karmann, Beust \& Klinger 2001, 2003, Thebault, Augereau \& Beust 2003). 

There are very substantial uncertainties regarding  the Oort belt comets in the solar
system.  Estimates of the total mass in this region range from   0.6 M$_{\oplus}$  (Stern \& Weissman 2001) to 40 M$_{\oplus}$ Weissman (1996), while there
are a  factor of 100 fewer observed ``dormant"  comets than  predicted with the simplest models (Levison et al. 2002).  

Given the major unknowns, in this paper
we describe  schematic rather than  detailed models.  We consider  the production of  gas-phase absorptions
in the observed spectra of main sequence stars caused by transiting comets which are warm enough that water-ice sublimates (see Beust, Karmann \& Lagrange 2001).     Previously, the effects of the photometric variations by cometary dust
have been computed by Lamers,  Lecavelier des Etangs \& Ferlet (1997) and Lecavelier des Etangs, Vidal-Madjar \& Ferlet (1999); here, we consider the possibility of detecting comets by their spectral line absorptions (see also Smith, Black \& Oppenheimer 1981).   We show results for  OH because it
is abundantly produced in comets  and because it can be detected from the ground.

\section{MODEL}
\subsection{Outflow from the Comet}

Following Whipple \& Huebner (1976) and Festou (1981), we adopt a simple  model for the gaseous outflow from  comets. When a comet approaches its host star, the ice is sublimated, and subsequently the
gas-phase H$_{2}$O is photodissociated mainly into OH and H:
\begin{equation}
h{\nu}\,+\,H_{2}O\;{\rightarrow}\;H\,+\,OH
\end{equation}
The resulting 
OH is mainly photodissociated into O and H:
\begin{equation}
h{\nu}\,+\,OH\;{\rightarrow}\,H\,+\,O
\end{equation}
 Finally, the O and H atoms are ionized, typically by charge exchange with the wind particles from the host star.  Below, we compute the spatial distribution of the OH molecules.    

We assume that the comet has an active area, $a$, and lies at a distance, $R$,
from the host star.  [We use capital and small letters to denote locations and speeds in
the coordinate system centered on the host star and comet, respectively.] We assume that the comet has a single surface temperature, and 
we focus on the parameter regime where the comet is sufficiently hot that cooling by ice sublimation balances heating by radiation from the host star and radiative cooling from the comet is relatively unimportant.  If ${\Delta}E$ is the average energy required to remove a water molecule from the comet, and $L_{*}$ denotes the luminosity of the host star, then the rate of production of H$_{2}$O molecules,
${\dot N_{H_{2}O}}$, is given by:
\begin{equation}
{\dot N_{H_{2}O}}\;=\;\frac{L_{*}\,a}{4\,{\pi}\,R^{2}\,{\Delta}E}
\end{equation}   
 The sublimation energy of pure water ice at 0 K is 7.6 ${\times}$ 10$^{-13}$ erg/molecule (for example, Ford \& Neufeld 2001).  However, not all of the  incident stellar energy is
converted into ice removal. We adopt ${\Delta}E$ = 2.0 ${\times}$ 10$^{-12}$ erg/molecule,  approximately the result given by Sekanina (2002) for the erosion energy
of Sun-grazing comets.  With equation (3) and our adopted value of ${\Delta}E$, we   reproduce within a factor of 2, the rates  of H$_{2}$O and OH production given by A'Hearn et al. (1995) 
and Crovisier et al. (2002) for ${\sim}$100  comets studied since 1976.    
For reference,  the comet with the largest measured outflow was Hale-Bopp where the  water molecule production rate at 1 AU was approximately 1.0 ${\times}$ 10$^{31}$ s$^{-1}$  (Dello Russo et al. 2000, Makinen et al. 2001, Harris et al. 2002, Morgenthaler et al. 2001 and
references), thus implying $a$ = 1.5 ${\times}$ 10$^{13}$ cm$^{2}$.   

To use equation (3), we assume that the comet is warm enough that sublimation is
the dominant cooling process.  This assumption fails for comet temperatures below ${\sim}$200 K, depending upon various physical parameters such as the infrared emissivity of the comet, the penetration of light into
the comet's nucleus and the exact composition of the comet (Whipple \& Huebner 1976).  Here, for  simplicity, we  assume that sublimation
from the comet follows equation (3) only if its distance from the host star is
less than an outer radius, $R_{out}$; otherwise we assume no sublimation. 
For numerical purposes,  we take $R_{out}$ = 2 AU.    

In writing equation (3), we assume that the sublimation rate from the comet does not depend upon whether it is inward-bound or outward-bound. This description
is appropriate for many Solar System comets (see Whipple \& Huebner 1976 and A'Hearn et al. 1995), and given the many other unknowns, this  seems like the appropriate approximation for the models presented here.  

Once the water is ejected from the comet, we assume a radial outflow
with a speed, $v_{O}$.   For most comets, 
 $v_{O}$ ${\approx}$ 1 km s$^{-1}$, but for comet Hale-Bopp, $v_{O}$ was approximately 3 km s$^{-1}$ since the gas was substantially heated as it
left the comet's nucleus (Harris et al. 2002). Because we are most interested in
the unusually massive comets like Hale-Bopp, we use $v_{O}$ = 3 km s$^{-1}$ in our
numerical examples.

  If $n_{H_{2}O}(r)$ denotes the density of water at distance, $r$, from the comet, then with spherical symmetry: 
\begin{equation}
\frac{1}{r^{2}}\frac{{\partial}(n_{H_{2}O}r^{2}v_{O})}{{\partial}r}\;=\;-J_{H_{2}O}n_{H_{2}O}
\end{equation}
where $J_{H_{2}O}(R)$ denotes the rate at which water is photodissociated.  
With the boundary condition determined by equation (3),  the solution to equation (4) is:
\begin{equation}
n_{H_{2}O}(r)\;=\;\frac{{\dot N_{H_{2}O}}}{4{\pi}r^{2}v_{O}}e^{-r/r_{H_{2}O}}
\end{equation}
In equation (5),  $r_{H_{2}O}$ = $v_{H_{2}O}/J_{H_{2}O}$, is the characteristic  distance which the water travels.  With $v_{O}$ = 3 km s$^{-1}$ and $J_{H_{2}O}$(1 AU) = 1.0 ${\times}$ 10$^{-5}$ s$^{-1}$ (Schleicher \& A'Hearn 1988, Harris et al. 2002), then $r_{H_{2}O}$ = 3.0 ${\times}$ 10$^{10}$ cm.    

If we define $n_{OH}(r)$ as the density of OH, then:
\begin{equation}
\frac{1}{r^{2}}\frac{{\partial}(n_{OH}r^{2}v_{O})}{{\partial}r}\;=\;-J_{OH}n_{OH}\;+\;J_{H_{2}O}n_{H_{2}O}
\end{equation}
By writing equation (6), we ignore the relatively small amount of water that
is photodissociated into O and H$_{2}$.  
We define $r_{OH}$ = $v_{O}/J_{OH}$, and
as long as $r_{OH}$ ${\neq}$ $r_{H_{2}O}$, 
 the solution to  equation (6) is:
\begin{equation}
n_{OH}(r)\;=\;\frac{{\dot N_{H_{2}O}}r_{OH}}{4{\pi}v_{O}\,(r_{H_{2}O}\,-\,r_{OH})\,r^{2}}\left(e^{-r/r_{H_{2}O}}\;-\;e^{-r/r_{OH}}\right)
\end{equation}
With   $v_{O}$ = 3 km s$^{-1}$ and $J_{OH}$(1 AU) = 7.5 ${\times}$ 10$^{-6}$ s$^{-1}$ (Schleicher \& A'Hearn 1988, Harris et al. 2002), then $r_{OH}$ = 4 ${\times}$ 10$^{10}$ cm. The OH photodissociation rate is a function of solar activity and can vary by a factor of 2 (Budzien, Festou \& Feldman 1994, van Dishoeck \& Dalgarno 1984). 

Close to the comet where both $r$ $<<$ $r_{H_{2}O}$ and $r$ $<<$ $r_{OH}$, we find from equation (7) that:
\begin{equation}
n_{OH}\;{\approx}\;\frac{{\dot N_{H_{2}O}}}{4{\pi}v_{O}r_{H_{2}O}r}
\end{equation}
The column 
density of OH, $N_{OH}$, along the line of sight with impact parameter $b$ relative to the comet's nucleus can be found with the assumptions 
that  $n_{OH}$ is given by equation (8) for  $r$ $<$ $r_{H_{2}O}$ and it is ``small" for
$r$ $>$ $r_{H_{2}O}$.  Thus, for $b$ $<<$ $r_{H_{2}O}$,  
\begin{equation}
N_{OH}\;{\approx}\;\frac{{\dot N_{H_{2}O}}}{2{\pi}v_{O}\,r_{H_{2}O}}\;\ln\frac{2\,r_{H_{2}O}}{b}
\end{equation}
Since ${\dot N_{H_{2}O}}$ varies as $R^{-2}$ and $r_{H_{2}O}$ varies as $R^{2}$, then $N_{OH}$ varies approximately as 
$R^{-4}$. Therefore, $N_{OH}$ is sensitive to the comet's orbital distance, $R$, from the host star.  

We show in Figure 1  the OH column density as
a function of displacement from an analog to Hale-Bopp  for an assumed distance from the Sun of 1 AU and $v_{O}$ = 3 km s$^{-1}$.  We consider both a case with  standard  ultraviolet dissociation rates, and  a case where $J_{OH}$ and $J_{H_{2}O}$ are increased by a factor of 10. In the standard case, the OH cloud is more extended and, for small values of $b$, the column density of OH is relatively low compared to the case with high ultraviolet intensity.   

Our results  for $N_{OH}$ allow us to determine where the OH is
optically thin.  As might be expected from pumping models for the populatons of the energy levels such as those
described by Schleicher \& A'Hearn (1988),  an OH energy level might  
contain at most 10\% of all the OH molecules.  Then, as seen in figure 1, even in the case of high ultraviolet radiation, its maximum column density is
less than ${\sim}$5 ${\times}$ 10$^{14}$ cm$^{-2}$.  Since the OH lines near 3100 {\AA} 
typically have oscillator strengths near   10$^{-3}$ (Roueff 1996), then  with $v_{O}$ = 3 km s$^{-1}$, these OH lines are optically thin. However, for $R$ $<$ 1 AU, the OH region is more compact
and the lines rapidly become  optically thick as $R$ decreases.

\subsection{The Comet's Orbital Motion}

The comet's orbit around the star, $R({\theta})$, is given by the expression:
\begin{equation}
R({\theta})\;=\;R_{p}\;\frac{1\,+\,{\epsilon}}{1\,+\,{\epsilon}\cos{\theta}}
\end{equation}
where $R_{p}$ is the distance at periastron, ${\epsilon}$ is the orbital eccentricity and ${\theta}$ is the angle measured between the major axis of the orbit's ellipse and the host star.
The comet's specific angular momentum, $L$, equals $R\,V_{\theta}$ where $V_{\theta}$ is the tangential speed of the comet.  If the mass of the star is $M_{*}$, we can write that:
\begin{equation}
L\;=\;([1\,+\,{\epsilon}]GM_{*}R_{p})^{1/2}
\end{equation}

\section{TRANSIENT OH ABSORPTION}
We now compute the absorption produced by the OH cloud around a comet.

\subsection{Threshold of Detectability}
 For the conditions of most interest, we expect that the size of the comet's OH cloud is smaller than the host star's radius.  In this case,  we only need to  compute the total number of
OH molecules produced in the outflow from a comet, ${\bf N(OH)}$. We may
use equation (7) to find  that:
\begin{equation}
{\bf N(OH)}\;=\;{\int}n_{OH}\,4{\pi}r^{2}dr\;=\;\frac{{\dot N_{H_{2}O}}}{J_{OH}}
\end{equation}
Equation (12) shows that  ${\bf N(OH)}$ is independent of the comet's distance to its host star because
both the water production rate (${\dot N_{H_{2}O}}$) and the OH photodissociation rate ($J_{OH}$) scale
as $R^{-2}$.  Equation (12) fails for $R$ $>$ $R_{out}$ where  water
is not appreciably sublimated.  
 
The average column density of OH absorbers, ${\overline N_{OH}}$ in the image of the host star, is given by the expression:
\begin{equation}
{\overline N_{OH}}\;=\;\frac{{\bf N(OH)}}{{\pi}R_{*}^{2}}
\end{equation}
If OH is optically thin in each portion of the stellar image and if for these
exploratory models  we ignore limb darkening (see Hubeny \& Heap 1996), then the total absorption in a spectral
line just depends upon ${\overline N_{OH}}$ and not the details on how the gas is spatially distributed.  Therefore, if
 the minimum column density for cometary  detection is $N_{min}$, then  using equations (3), (12) and (13), we require  that:
\begin{equation}
a\;{\geq}\;\frac{4{\pi}^{2}{\Delta}E\,J_{OH}(R)R^{2}R_{*}^{2}N_{min}}{L_{*}}
\end{equation}
Since $J_{OH}$ varies as $R^{-2}$,  we see from equation (14) that $a$ is independent of $R$.

The value of $N_{min}$ depends upon the  telescopes and instruments that are used. 
Within the interstellar medium where the gas has a similar velocity dispersion to what we expect for the outflow from a comet, OH column densities of ${\sim}$10$^{13}$ cm$^{-2}$ have been measured (Roueff 1996).  Although the OH produced by comets is distributed over more energy levels than the OH within the interstellar medium (see, for example, Schleicher \& A'Hearn 1988), since the same number of photons
are absorbed from the background source if the OH is optically thin, it may be possible with modern detectors and cross
correlation techniques  to detect total OH column densities in  cometary outflows as
low as 10$^{13}$ cm$^{-2}$.    
If  $N_{min}$ = 10$^{13}$ cm$^{-2}$, we find for environments similar to the solar system's that $a$ ${\geq}$ 170 km$^{2}$; analogs to comet Hale-Bopp
with $a$ = 1500 km$^{2}$ easily satisfy this criterion.  Moreover, from equation (14), if $N_{Min}$ = 10$^{13}$ cm$^{-2}$, then an analog to comet Hale-Bopp could be detected around a G-type star where the ultraviolet luminosity is as much as a factor of 9 larger than the Sun's.

\subsection{Likelihood of an Absorption Event}
 
In our simple model, a comet is only active if $R$ $<$ $R_{out}$. From equation (10), we find that for the half of the orbit between ${\theta}$ = 0 and
${\theta}$ = ${\pi}$, the maximum angle for cometary activity, ${\theta}_{max}$, is: 
\begin{equation}
{\theta}_{max}\;=\;\cos^{-1}\left(\frac{1\,+\,{\epsilon}}{\epsilon}\,\frac{R_{p}}{R_{out}}\,-\frac{1}{{\epsilon}}\right)
\end{equation} 

For simplicity, we assume that the comet's outflow is undetectable
for $R$ $<$ $R_{in}$ where $R_{in}$ is defined as the region where the OH lines are optically thick.  We introduce a minimum angle, ${\theta}_{min}$, in a fashion similar to our definition
of ${\theta}_{max}$ in equation (15) except that we substitute $R_{in}$ for $R_{out}$.  As described below, we are most interested in environments where
the ultraviolet luminosity of the host star is much larger than the Sun's.
Therefore, we take
 $R_{in}$ = 1 AU since, as shown in Figure 1,  this is the region where the OH column density exceeds 10$^{14}$ cm$^{-2}$ and therefore the outflow from the comet becomes compact and opaque.
  With $R_{in}$ = 1 AU, $R_{out}$ = 2 AU and  ${\epsilon}$ ${\approx}$ 1, we list in Table 1 values of ${\theta}_{min}$ and  ${\theta}_{max}$ for analogs to those     solar system comets with  $a$ $>$ 50 km$^{-2}$ and $R_{p}$ $<$ 2 AU  from A'Hearn et al. (1995) and  also comet Hale-Bopp.

In order for us to detect transient absorption, the active comet must  transit the host star when $R_{in}$ $<$ $R$ $<$ $R_{out}$.  From the perspective of the host star, we imagine two strips of solid angle, one for the inbound and one for the outbound portions of the comet's orbit, through which the comet must pass.  In each strip, an element of solid angle where the comet transits the host star is defined by the angular height, $(2\,R_{*}/R)$, multiplied by the angular width, $d{\theta}$.  Thus, the probability that we can witness transient absorption, $p$, is given by the fraction of sky covered by
the two strips,  
or, using equation (10):
\begin{equation}
p\;{\approx}\;\frac{2}{4{\pi}}{\int}_{{\theta}_{min}}^{{\theta}_{max}}\frac{2\,R_{*}}{R}\,d{\theta}\;=\;\frac{R_{*}([{\theta}_{max}\,-\,{\theta}_{min}]\,+\,{\epsilon}[\sin{\theta}_{max}\,-\,\sin{\theta}_{min}])}{{\pi}R_{p}(1\,+\,{\epsilon})}
\end{equation}
We list in Table 1 values of $p$ for analogs to recently observed comets; $p$  typically is near 0.1\%.

\subsection{Duration and Probability of  Absorption Events}

Above, we have computed the probability that sometime during its orbit
a comet can be detected. We now estimate the duration, $t_{abs}$, of each absorption event.     We take for an ``average" cometary transit time that
$t_{abs}$ =   ${\sqrt{2}}R_{*}/V_{\theta}$.  We then write: 
\begin{equation}
t_{abs}\;=\;\frac{{\sqrt{2}}\,R_{*}\,R}{L}
\end{equation}
Combining equations (10), (11) and (17), we then find that:

\begin{equation}
t_{abs}({\theta})\;=\;\frac{{\sqrt{2}}\,R_{*}\,R_{p}^{1/2}(1\,+\,{\epsilon})^{1/2}}{(1\,+\,{\epsilon}\cos{\theta})(GM_{*})^{1/2}}
\end{equation}
For ${\epsilon}$ ${\sim}$ 1, ${\theta}$ $=$ ${\pi}/2$, $R_{p}$ = 1 AU, and $M$ = 1 M$_{\odot}$ star, the characteristic duration of the absorption event is  5${\times}$ 10$^{4}$ s.  

The total probability, $P(OH)$, of an observation leading to a detectable transient absorption is given by the expression:
\begin{equation}
P(OH)\; = \;p\,t_{abs}\,{\dot T}
\end{equation} 
where ${\dot T}$ (s$^{-1}$) is the rate at  which sufficiently large comets  arrive with $R_{p}$ $<$ $R_{out}$. 
Numerically, with ${\epsilon}$ ${\approx}$ 1, ${\theta}$ = ${\pi}/2$, $R_{p}$ = 1 AU, $M$ = 1 M$_{\odot}$ and
$p$ = 10$^{-3}$, then $P(OH)$ = 47 ${\dot T}$.

Even for the solar system, 
 ${\dot T}$ for the large comets that are most likely to be detected  is uncertain.
A'Hearn et al. (1995) identified 4  comets with  $a$ $>$ 50 km$^{-2}$ and $R_{p}$ $<$ 1 AU  during 16 years.  For these objects, then ${\dot T}$ = 8 ${\times}$ 10$^{-9}$ s$^{-1}$.
In the solar system, the cumulative number of comets with area larger than $a$ varies approximately as $a^{-0.7}$, at least for objects with radii between
1 km and 10 km.
(See Meech, Hainaut \& Marsden 2004 and convert from radius distribution to
area distribution).    Therefore, scaling from $a$ = 50 km$^{-2}$ to $a$ = 1500 km$^{-2}$,  
 we  expect that comets at least as large as Hale-Bopp  have an arrival rate in the inner solar system of
7 ${\times}$ 10$^{-10}$ s$^{-1}$. This estimate implies that a very large comet like Hale-Bopp should arrive every 50 years, a rate that  is approximately consistent with the historic data.  For example,   
  Sekanina (2002) has
argued that comet 1882 R1 had a size similar to Hale-Bopp's.    With  ${\dot T}$ = 7 ${\times}$ 10$^{-10}$ s$^{-1}$,  we find from equation (19) that $P(OH)$ = 3 ${\times}$ 10$^{-8}$.  Thus, the probability of detecting an individual comet around an analog to the Sun is very low.  

\section{COMETS AROUND YOUNG STARS}

Above, we have considered the possibility of detecting comets around stars
similar to the Sun.  Here, we consider extrapolating these models to
models for the Sun at earlier ages.  There are two competing
factors. First, the younger stars are more active and may emit as much as 50 times more
ultraviolet light (Ayres 1997, Guinan, Ribas \& Harper 2003).  However, the stellar ultraviolet
luminosity and level of activity is variable, and some young stars have
ultraviolet luminosities ``only" ${\sim}$10 times the solar value.  Thus
analogs to comet Hale-Bopp might be detectable.           
   Second, there are both theoretical and observational reasons for thinking that young solar-type stars might have
vastly greater comet arrival rates than is currently witnessed in the solar
system.  Models for the formation of the Earth all invoke the build-up of
smaller planetesimals.  
In this context, Wetherill (1992) has computed that in young planetary systems where there is no analog to Jupiter, the
rate of cometary periastron passages less than 1 AU could be 3 ${\times}$ 10$^{5}$ greater
than is currently seen in the solar system.   Also, models for the  delivery of water to the Earth describe how giant planets perturb the orbits
of icy planetesimals into elliptical orbits which result in matter from the outer
planetary system being injected to within 1 AU of  the host star (Morbidelli et al. 2000).  These planetesimals may be asteroids (Petit, Morbidelli \& Chambers 2001) or comets.  
To estimate the magnitude of this effect, consider that the total mass of the Earth's ocean is 1.4 ${\times}$ 10$^{24}$ g.  If this
water was mainly delivered during the Earth's first 50 Myr, then the water accretion
rate onto the Earth was  ${\sim}$10$^{9}$ g s$^{-1}$. Currently, the zodiacal light is explained by a dust infall rate of 3 ${\times}$ 10$^{6}$ g s$^{-1}$ (Fixsen \& Dwek 2002) while the Earth
only  accretes  about 10$^{3}$ g s$^{-1}$ of dust (Love \& Brownlee 1993).  (While dust in the ecliptic plane largely results from collisions between asteroids [Grogan, Dermott \& Durda 2001], when considering the entire sky, an appreciable fraction of the  current zodiacal cloud results from the disintegration of comets [Hahn et al. 2002]).  During the era of ocean formation,  if the efficiency    of the accretion by the Earth of dust with a comet-like origin was 3 ${\times}$ 10$^{-4}$ as suggested by the current
environment, then the inner solar
system may have experienced a total infall rate of water-rich planetesimals of ${\sim}$10$^{12}$ g s$^{-1}$.      Therefore, it is  possible that in the early solar
system, the rate of arrival of icy objects within 1 AU was ${\sim}$3 ${\times}$ 10$^{5}$ times
larger than the current rate of the arrival of comets. This enhancement of the  cometary arrival
rate suggests that the probability of any single OH-sensitive observation 
detecting a large Hale-Bopp analog in such a young system approaches 1\%.

\section{DISCUSSION}

Above, we have discussed models for the detectability of comets.  Here,
we suggest possible targets for an observational program.
Although many observers avoid working in the near ultraviolet,  Israelian, Garcia Lopez \& Rebolo (1998) and Boesgaard et al. (1999) have observed the  OH bands near 3100 {\AA} in solar-type stars to infer their oxygen abundances.  It may be possible to extend
such studies to search for comets.

Comets within 2 AU of a solar-type star not only sublimate ice, but they also eject dust into their surroundings. This dust
has a characteristic temperature  $>$200 K which results in substantial emission at
25 ${\mu}$m.  Previous studies with IRAS and ISO of dust around solar-type stars have not revealed much infrared excess at ${\lambda}$ $<$ 25 ${\mu}$m (Laureijs et al. 2002, Zuckerman \& Song 2004).  However, the enormous increase in sensitivity provided by  {\it Spitzer} (Werner et al. 2004) has allowed the study of circumstellar grains around  many more solar-type stars.  Specifically, recent {\it Spitzer} observations of 15 Myr old solar-type stars in the Lower Centaurus Crux Association show substantial  24 ${\mu}$m excesses where  the dust production rates  may be 10$^{6}$ greater than the dust production rate for the zodiacal light
within the solar system (Chen et al. 2004).    These same stars also display
relatively low X-ray luminosities compared to other solar-type stars in the
same Association,  and consequently, they may also have relatively low
ultraviolet luminosities. Therefore they may be suitable targets
for  comet-searching. Some solar-type stars as old as 100 Myr also show
substantial 24 ${\micron}$ excesses (Gorlova et al. 2004).    
 
Unfortunately,
the prospects for detecting transient OH absorption around  warmer stars like ${\beta}$ Pic ($L$ = 8.5 L$_{\odot}$; $T_{eff}$ = 8200 K; Crifo et al. 1997)    are not very good. That is, ${\beta}$ Pic emits enough photospheric light between 1600 {\AA} and 1800 {\AA}, where both OH and water have significant photodissociation cross sections, that the lifetimes of these molecules are
relatively short.  In particular, 
given the star's luminosity, we would expect that
comets at 3 AU around ${\beta}$ Pic might resemble comets at 1 AU around the Sun. Using archived HST data (see Roberge et al. 2000),  scaling the observed ultraviolet flux upwards by (19.3 pc/3 AU)$^{2}$, and using the compilation of
the photodestruction cross sections by Budzien et al. (1994), we find that the  
 lifetimes at 3 AU around ${\beta}$ Pic of OH and H$_{2}$O are ${\sim}$1600 s and ${\sim}$330 s, respectively.  
As a result, from equation (14), only comets with areas greater than 3 ${\times}$ 10$^{4}$ km$^{2}$ would be large enough to produce an OH column density of 10$^{13}$ cm$^{-2}$.  Thus transient OH absorption spectrum is expected to be very weak unless the comet is vastly larger than Hale-Bopp. Some  gas-phase CO is seen in the ultraviolet absorption-line spectrum of ${\beta}$ Pic, but it does not change with time and it is probably not directly associated with infalling  comets (Roberge et al. 2000).

\section{CONCLUSIONS}

We have considered the transient OH absorption signature of comets similar to those found in the solar system.  Such events are very rare for stars like the Sun. For young solar-type stars with a substantial 24 ${\mu}$m excess  where there may be comets  delivering water to terrestrial planets,  a single, sensitive, randomly-timed observation may have a 1\% chance of detecting OH flowing away from a large comet. 

This work has been partly supported by NASA.

\newpage
\begin{center}
Table 1  -- OH Detectability of Analogs to Recent Bright Comets

\begin{tabular}{lrrrrl}
\hline
\hline
Comet & $R_{p}$  & $a$ &  ${\theta}_{min}$ &${\theta}_{max}$ &    $p$ \\
   & (AU) &  (km$^{2}$) &  ($^{\circ}$) &   ($^{\circ}$) \\
\hline
Bradfield & 0.26  & 96 & 119 & 138 &0.00036 \\
Halley & 0.59  & 50 & 80 & 114 & 0.00066\\
Levy & 0.94  & 110 & 28&93& 0.0013\\
Meier & 1.14  & 250& 0 & 82& 0.0016 \\
Wilson & 1.20  & 79 & 0 &78 & 0.0015\\
West & 0.20  & 130 & 127 & 143&0.00031\\
Hale-Bopp & 0.91  & 1500 & 35 & 95 & 0.0012 \\
\hline
\end{tabular}
\end{center}
We take ${\theta}_{max}$ and $p$ from equations (15) and (16), respectively.
The cometary parameters are from A'Hearn et al. (1995) except for Hale-Bopp which is discussed in the text.    
\newpage
\begin{figure}
\epsscale{1}
\plotone{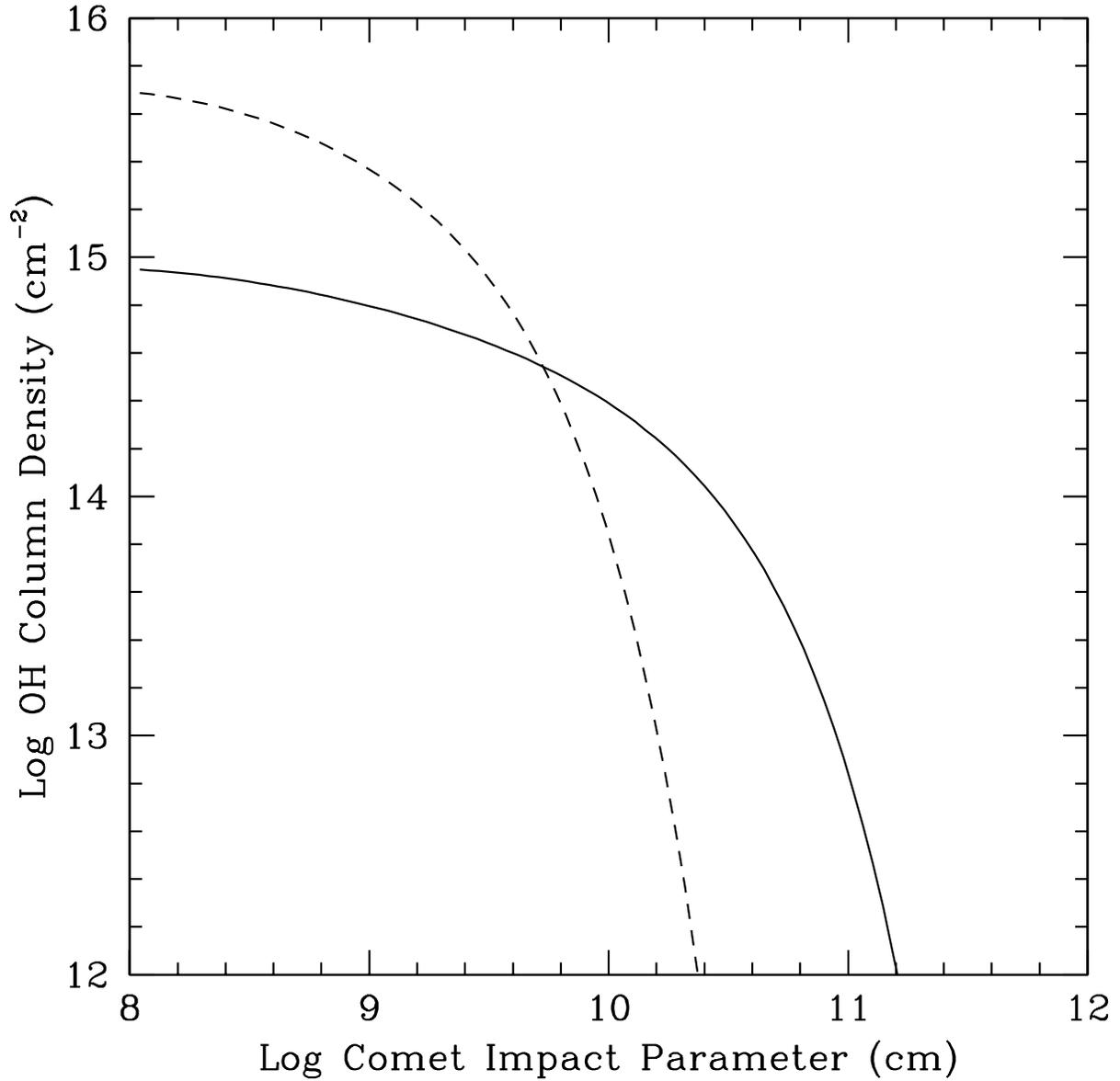}
\caption{Plot of OH column density vs. impact parameter  for an analog to comet Hale-Bopp with $a$ = 1500 km$^{2}$ located at 1 AU from a star with $L$ = 1 L$_{\odot}$.  We use $v_{O}$ = 3 km s$^{-1}$ both for the case with standard  photodissociation rates (solid line) and for the case where these
 rates are increased by a factor of 10 (dashed line).}
\end{figure}
\end{document}